\documentclass[prl,superscriptaddress,showpacs,twocolumn]{revtex4}
\usepackage{srcltx}
\usepackage{amssymb,amsmath,amsfonts,graphicx,hyperref}%

\begin{document}
\title{Compaction in Granular Solid Hydrodynamics}
\author{Yimin Jiang}
\affiliation{Central South University, Changsha 410083,
China}
\author{Mario Liu}
\affiliation{Theoretische Physik, Universit\"{a}t
T\"{u}bingen,72076 T\"{u}bingen, Germany}
\date{\today}
\begin{abstract}
Compaction is considered and embedded into broader
granular behavior. Reversible compaction is related to
the pressure exerted by agitated grains, a quantity
relevant to dense flow. Irreversible compaction is
derived from the loss of elastic deformation, the
physics behind elasto-plastic flows.
\end{abstract}  \maketitle

Granular compaction -- density increase by gentle
perturbation of the grains -- is a ubiquitous
phenomenon, relevant in many contexts such as
efficient packaging, or when establishing the
structural stability of off-shore windmills that are
subject to continual surfs. Compaction has been
studied widely but is not yet fully understood, see
the review~\cite{1}, which carefully considers the
phenomenon and cites many useful references. There are
two compaction branches: In the irreversible one, the
density increases monotonically; in the reversible
one, one observes compaction and loosening, depending
on whether the perturbation is weakened or
strengthened.

Granular solid hydrodynamics ({\sc gsh}) is a set of
partial differential equations~\cite{JL6} constructed
to model granular medium in all its facets, though
qualitatively at present. It was first employed to
calculate static stress distribution for various
geometries, including sand piles, silos, and point
load, achieving results in agreement with
observation~\cite{ge1}. It was then employed to
consider slowly strained granular solid~\cite{JL4},
and found to yield response envelopes similar to those
from the hypoplastic theory~\cite{Kolym-1}. Recently,
the critical state -- generally considered a hallmark
of granular behavior~\cite{CSSM} -- was identified as
a steady-state, elastic solution of {\sc
gsh}~\cite{critState}. Uniform dense flow,
fluidization and jamming were also
considered~\cite{denseFlow}, finding broad agreement
to experiments, and more narrowly focused existing
theories~\cite{chute2,Lub-1}. Finally, the velocity of
elastic waves were calculated as a function of the
stress~\cite{elaWave}.

Here, we use {\sc gsh} to consider compaction. At
present, ``geometric frustration models"~\cite{1} are
employed to account for the irreversible branch, and
the Edward entropy $S_{Ed}$~\cite{Edw} for the
reversible one. Both are again narrowly focused.
Moreover, the notion of $S_{Ed}$ entails hidden
assumptions that merit closer scrutiny.

Derived in compliance with general principles, {\sc
gsh} consists of \textbullet~five conservation laws
for the density respectively of the energy $w$, mass
$\rho$, and momentum $\rho v_i$, \textbullet~an
evolution equation for the elastic strain $u_{ij}$,
and \textbullet~balance equations for two entropy
densities, the true and the granular one, $s$ and
$s_g$. Two entropies are necessary because of granular
media's {\em two-stage irreversibility}: Macroscopic
energy, kinetic and elastic, dissipates into
mesoscopic, inter-granular degrees of freedom, mainly
granular jiggling and the collision-induced,
fluctuating elastic deformation. After a
characteristic time, the energy degrades further into
microscopic, inner-granular degrees of freedom,
especially phonons. The two entropies account
respectively for the energy of the meso- and
microscopic degrees of freedom.

Now a brief presentation of the  {\sc gsh}-equations,
tailored for the purpose of granular compaction.
(See~\cite{JL6} for the complete theory.) An important
quantity of {\sc gsh} is the energy density $w$, that
in its local rest frame is a function of the above
listed four variables, $w_0(\rho,u_{ij},s_g,s)$. The
conjugate variables: true temperature
$T\equiv{\partial w_0}/{\partial s}$, elastic stress
$\pi_{ij}\equiv$ $-{\partial w_0}/{\partial u_{ij}}$,
chemical potential $\mu\equiv{\partial
w_0}/{\partial\rho}$, granular temperature
$T_g\equiv{\partial w_0}/{\partial s_g}$, and gaseous
pressure $P_T\equiv$ $\mu\rho+s_gT_g+sT-w_0$, are
given once $w_0$ is. The Cauchy stress is given as
$(1-\alpha) \pi_{ij}+P_T\delta_{ij}$, [$\alpha$ is
explained below Eq~(\ref{gc2}),] of which only the
trace is needed here,
\begin{equation}\label{gc3}
P=(1-\alpha)\pi_{\ell\ell}/3+P_T.
\end{equation}
In {\sc gsh}, the elastic strain $u_{ij}$ is taken as
the portion of the strain that deforms the grains and
leads to reversible storage of elastic energy. The
energetically indifferent rest, such as rolling and
sliding, is the plastic portion. Again, only
$\Delta\equiv-u_{\ell\ell}$ is relevant. Its
evolution, taken as
\begin{equation}\label{gc2}
\partial_t \Delta+(1-\alpha)
{v}_{\ell\ell}=-\Delta/\tau,\quad 1/\tau=\lambda T_g,
\end{equation}
accounts for two facts: First, only the portion
$1-\alpha$ of the strain rate $v_{ij}\equiv\frac12
(\nabla_i{v}_j+\nabla_j{v}_i)$ deforms the grains,
changing $\Delta$ and the energy $w_0(\Delta)$ -- the
rest is plastic. That the same $\alpha$ appears in
Eq~(\ref{gc3}) is a result of the Onsager relation.
But it may also be understood via an analogy to a
bicycle gear shift: A bigger $\alpha$ implies more
pedaling at reduced torque, both with the same factor.
Second, when grains jiggle and $T_g$ is finite, the
grains briefly lose contact with one another. The
result is a slow relaxation of granular deformation --
as expressed by $\Delta/\tau$. The full tensorial
version of Eq~(\ref{gc2}) is closely related to
hypoplasticity, and yields a realistic account of the
elasto-plastic dynamics of  granular media,
see~\cite{JL4}.

Because Eqs~(\ref{gc3},\ref{gc2}) are fairly general,
it is the energy density $w$ that encodes the
material-specific information. The expression we
consistently employed to study static stress
distribution and hypoplastic deformation is: $w_0$
$=w_1(u_{ij},\rho)$ $ +w_2(s_g,\rho)+w_3(s,\rho)$,
where $w_1$ is the macroscopic, $w_2$ the mesoscopic,
and $w_3$ the microscopic contributions. We take them
(if without shear strain and away from the virgin
consolidation line) as
\begin{equation}\label{gc4}
w_1(u_{ij},\rho)={\mathcal B}\Delta^{2.5}/2.5, \quad
w_2(s_g,\rho)=s_g^2/(2\rho b),
\end{equation}
and  $w_3=e\rho/m$. Explanations: \textbullet~The
elastic energy $w_1$ contains the lowest order term in
$\Delta$. Being 2.5 rather than two, the exponent
accounts for the Hertz-like contacts between the
grains and the Coulomb yield~\cite{JL1}. As granular
systems are slightly stiffer at higher densities, the
elastic coefficient ${\cal B}$
grows slowly and monotonically with $\rho$. In the
present context, however, we may approximate ${\cal
B}$ as a constant -- of around 8 GPa for river sand
and 7 GPa for glass beads~\cite{ge1} -- because the
density dependence of $w_2$ is the dominant one.
\textbullet~$w_2$ contains the lowest order term in an
expansion in $s_g$. The exponent is two because $w_2$
has a minimum at $s_g=0$. The density dependence of
$b(\rho)=b_0(1-\rho/\rho_{cp})^{a}$, with $a=0.1$, is
chosen to reproduce the observed, volume-dilating
pressure contribution from agitated grains, $P_T\sim
w_2/(\rho_{cp}-\rho)$, that diverges at the random
close packing density $\rho_{cp}$, see~\cite{Lub-1}.
The value for the constant $b_0$ depends on the scale
we choose for $T_g$. Requiring (somewhat arbitrarily
but adhering to convention) that $\frac32k_BT_g$ is,
in the rarefied limit of granular gas,  the kinetic
energy of a single grain, we find
$b_0\approx10^{-25}{\rm J/(KgKelvin^2)}$.
\textbullet~Finally, $e,m$ are respectively the
average internal energy and mass per grain. Being
linear in $\rho$, $w_3$ does not contribute to $P_T$
and is not relevant for compaction. Given
Eq~(\ref{gc4}), the associated pressure may be
calculated from Eq~(\ref{gc3}) as
\begin{equation}\label{gc5}
P={(1-\alpha)}{\cal B}\Delta^{1.5}+{a\rho^2 b\,
T_g^2}/{[2(\rho_{cp}-\rho)]}.
\end{equation}
Eqs~(\ref{gc2},\ref{gc5}) are all one needs to account
for both branches of compaction. This ends the
presentation of {\sc gsh}.

Assuming constant $T_g,P$ while evaluating
Eqs~(\ref{gc2},\ref{gc5}) simplifies the calculation
and focuses the attention on the universal aspects of
granular compaction. The underlying physics is: At
given $T_g$, the elastic deformation $\Delta$ relaxes
according to Eq~(\ref{gc2}). To maintain the constant
pressure $P$, the density must increase as prescribed
by Eq~(\ref{gc5}). This is the irreversible branch of
compaction. Relaxation stops at $\Delta=0$, when the
pressure is purely temperature generated,
$P=P_T={a\rho^2 b\, T_g^2}/{[2(\rho_{cp}-\rho)]}$, and
the density assumes the final value $\rho_f$,
\begin{eqnarray}\label{gc6}
2\rho _{cp}/\rho _{f}=1+\sqrt{1+2ab_{0}\rho
_{cp}T_{g}^{2}/P},
\end{eqnarray}
obtained by taking $b=$const. Modifying the magnitude
of perturbation now,  $\rho_f$ will adjust -- becoming
larger the smaller $T_g$ is. This further density
modification is of thermodynamic origin and
reversible.

Changing $T_g$ midway, with $\Delta$ still finite,
leads to a change in $\Delta$ as well, if $P$ of
Eq~(\ref{gc5}) is given. This disrupts the relaxation
of $\Delta$, in essence resetting its initial
condition -- as observed in~\cite{mem} and taken as a
memory effect. (Generally speaking, ``memory" is
usually a result of hidden variables: When all
variables have the same values, but the system still
behaves differently, we speak of memory, or
history-dependence. But an overlooked variable that
has different values for the two cases will naturally
explain the difference. The manifest and hidden
variables here, clearly, are $\rho$ and $\Delta$,
respectively.)

$T_g$ is of course not always constant, but neither
does it have to be. It is constant for constant shear,
or sound waves that oscillates faster than $T_g$
relaxes; it is not for tapping or sinusoidal shear,
which give rise to periodic flare-ups of $T_g$. Yet
because compaction is the result of tiny plastic
deformation accumulated over many periods, it is the
temperature $\langle T_g\rangle$, averaged over many
periods, that is relevant. On this longer time scale,
constant $T_g$ is an appropriate approximation. Also,
the pressure $P$ is frequently nonuniform, such as
when grains are filled in an open vessel, though it is
still given. Being local expressions, all above
equations remain valid, and we can read off say
$\rho_f(\boldsymbol r)$ from Eq~(\ref{gc6}) if $P(r)$
is known.

In a complete theory, one needs to go beyond the
universal aspects of compaction, and account for how
$T_g$ is excited by the given perturbation and related
to its amplitude. We shall not do this here, only note
that one can in principle achieve this by solving {\sc
gsh} for given boundary conditions. These include
oscillatory shear, intense sound waves, periodic
liquid injections, and horizontal tapping in various
geometries, but unfortunately not vertical tapping,
because particles flying ballistically part of the
time is not a phenomenon amenable to any hydrodynamic
description.

The simplest perturbation to account for is
oscillatory shear. As long as the yield surface is not
breached, and the frequency of oscillation small
compared to the relaxation rate of $T_g$, one obtains
(from the balance equation for $s_g$) a quasi linear
relationship between $T_g$ and the strain rate,
$T_g\sim\sqrt{v_{ij}v_{ij}}$, see~\cite{JL6,JL4}.
Liquid injection is similar but involves spatial
variation. A stationary sound field, if uniform, is
again simple, as it maintains a constant $T_g$ via the
strain rate that accompanies any sound. When the sound
is turned off, $T_g$ is quickly gone. This increases
$\Delta$ to maintain the pressure, and {\em quenches
the density at the value characteristic of $T_g$.}
This is why we may stop the perturbation to measure
the density. (During the quench, there is an elastic
density change that is negligibly small, but may be
accounted for anyway.)

If a sand-filled vessel is tapped horizontally, the
blow generates a pulse of sound waves. As the pulse
propagates through the medium and reflects multiply
off the wall, it becomes a more or less uniform sound
field -- of brief duration but still with an
accompanying $T_g$ that is monotonically related to
the strength of the blow. And the quenched density
measured after the tap is the one characteristic of
this $T_g$. All these may in principle be evaluated
employing {\sc gsh}. Tapping the vessel vertically,
circumstances are somewhat different, as a fraction of
the particles will fly ballistically some of the time.
During the flight, the grains will keep their kinetic
energy, but gradually lose $\Delta$: When the squeeze
from the neighboring grains is suddenly gone after the
tap, each grain oscillates around its free shape with
a diminishing amplitude. This is puzzling, since
compaction seems a succession of brief
$\Delta$-relaxation, each picking up the work where
the last one left it, and it is unclear why this
process is not disrupted by $\Delta$ vanishing
repeatedly. A conceivable answer is that these taps
are not that strong.

Newer evidence~\cite{hecke} shows that the control
parameter of the reversible branch is not the
acceleration $\Gamma$, but the initial velocity
$t\Gamma$, with $t$ the tap duration. This is consistent
with {\sc gsh}, because with $w_2\sim s_g^2\sim T_g^2\sim
(t\Gamma)^2$, Eq~(\ref{gc6}) indeed gives $\rho_f$ as a
function of $t\Gamma$.

Eqs~(\ref{gc2},\ref{gc5}) are now evaluated with
$T_g,P$ taken as constants. Starting from
Eq~(\ref{gc2}) and the continuity equation,
$\partial_t\rho/\rho=- v_{\ell\ell}$, we obtain
$({1-\alpha}){\partial_t\rho}/{\rho}={\partial_t\Delta+\Delta/\tau}$.
From this, we eliminate either $\partial_t\Delta$ or
$\partial_t\rho$ by employing $\partial_t
P(\Delta,\rho)=0$, arriving respectively at
\begin{equation}\label{gc9}
\partial_t\Delta=-\Delta/\tau_\Delta,\quad
\partial_t\rho=\rho/\tau_\rho,
\end{equation}
with $\tau_\Delta\equiv{\tau}[ 1+{(1-\alpha)}/{\rho A
}]$, $\tau_\rho\equiv({\tau}/\Delta) [1-\alpha+\rho
A]$, and $A\equiv({\partial P}/{\partial\rho})/
({\partial P}/{\partial\Delta})=-\left.
{\partial\Delta}/{\partial\rho}\right|_{P}$.
(Employing Eq~(\ref{gc5}), both $\tau_\rho$ and
$\tau_\Delta$ may be written as a function of either
$\Delta$ or $\rho$ alone.) Note $\tau_\rho,
\tau_\Delta$ are not proper relaxation times, as they
depend on $\Delta,\rho$. Nevertheless, we see that
$\Delta$ diminishes, while $\rho$ grows -- or relaxes
backward in time. Since $\partial
P/\partial\Delta\sim\sqrt\Delta$, or
$\tau_\Delta\to\tau$ for $\Delta\to0$, $\Delta$ will
eventually relax exponentially toward zero,
characterized by the constant relaxation rate
$1/\tau=\lambda T_g$ for $T_g=$ constant. At the same
time, $\rho$ grows, ever more slowly, with the
diminishing rate $1/\tau_\rho\sim\Delta^{1.5}/{\tau}$.
It stops completely at $\rho=\rho_f$ for $\Delta=0$.
Integrating Eqs~(\ref{gc9}) yields $\rho(t)$,
accounting for irreversible compaction. The reversible
branch corresponds to the change of $\rho_f$ with
$T_g$ as given by Eq~(\ref{gc6}). As this relation is
thermodynamic in nature, it holds independent of path.

We conclude that {\sc gsh} is capable of a
transparent, de-mystified account for the universal
aspects of compaction. It describes slow density
growth in the presence of granular jiggling, with a
characteristic time that diverges towards the end.
This happens concurrent with the lost of elastic
deformation. The final density, a thermodynamic
function of $T_g$, is larger the smaller $T_g$ is. Two
things still need to be done: First, since a crucial
starting point of {\sc gsh} is the second law of
thermodynamics using a conventional entropy, we need
to understand its relation to the Edwards entropy.
Second, we shall make contact with experiments, to
find agreement that are, in spite of  {\sc gsh}'s
qualitative status, satisfactory.

We revisit {\it granular statistical mechanics} and
the Edwards entropy $S_{Ed}$. Substituting the volume
$V$ for the energy $E$, and compactivity $X$ for the
temperature $T$, this theory takes ${\rm d}V=X{\rm
d}S_{Ed}$ as the basic thermodynamic relation for a
{\it``mechanically stable agglomerate of infinitely
rigid grains at rest"}~\cite{Edw}. The entropy
$S_{Ed}$ is obtained by counting the number of
possibilities to package grains for a given volume,
equating it to $e^{S_{Ed}}$. Because a stable
agglomerate is stuck in one single configuration, some
perturbation is necessary to enable the system to
explore the phase space of $S_{Ed}$. This {\em
ansatz}, we believe, is best appreciated by taking the
entropy more generally as $S(E,V)$, or ${\rm
d}S=(1/T){\rm d}E+(P/T){\rm d}V$, with
$1/T\equiv\partial S/\partial E$, $P/T\equiv\partial
S/\partial V$. For infinitely rigid grains at rest, we
have $E\equiv0$, because the energy remains zero
however these non-interacting grains are arranged.
Therefore, ${\rm d}S=(P/T){\rm d}V$, or ${\rm
d}V=(T/P){\rm d}S\equiv X{\rm d}S$.

In {\sc gsh}, grains are neither assumed infinitely
rigid, nor always at rest. The energy, containing both
elastic and kinetic contributions, does not vanish --
though it of course does for infinitely rigid grains
at rest, for $u_{ij},v_i,T_g\equiv0$. Therefore,
$S_{Ed}$ is a minimalistic approach, a special limit
of {\sc gsh}. Note, however, the following objections:
\textbullet~Because of the Hertz-like contact between
grains, very little material is being deformed at
first contact, and the compressibility diverges at
vanishing compression. This is a geometric fact
independent of how rigid the bulk material is.
Infinite rigidity is never a realistic limit for sand.
\textbullet~The number of possibilities to arrange
grains for a given volume concerns inter-granular
degrees of freedom. These are vastly overwhelmed by
the much more numerous configurations of the
inner-granular degrees of freedom. Maximal entropy $S$
implies minimal macroscopic and mesoscopic energy. It
is quite unrelated to maximal number of possibilities
to package grains. \textbullet~Being an equilibrium
consideration, $S_{Ed}$ is meant to account for the
reversible branch of compaction. Yet Eq~(\ref{gc6}) is
a consequence of the pressure exerted by agitated
grains, not grains at rest. \textbullet~Granular media
have two different equilibria, a solid-like one for
$T_g=0$, when the system is jammed, and a liquid-like
one for finite $T_g$, see III of \cite{JL6}. A
relaxing $\Delta$ represents an advance toward the
liquid equilibrium, not reflected in $S_{Ed}$. For all
these reasons, we are skeptical that $S_{Ed}$, though
clearly a simplification, represents a useful limit.

\begin{figure}[b]
\begin{center}
\includegraphics[scale=.9]{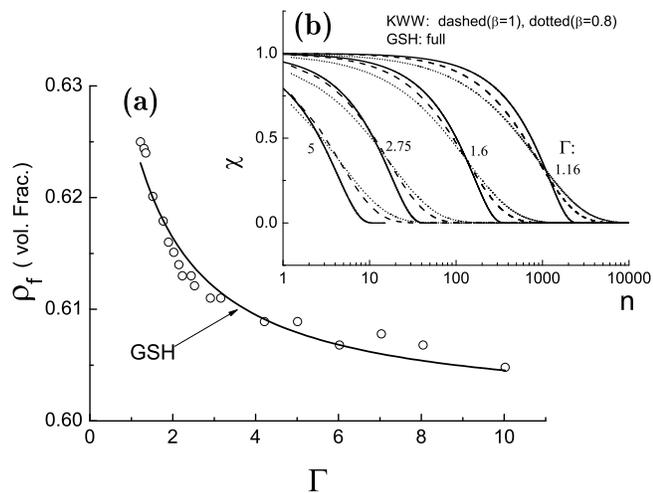}
\end{center}
\caption{Comparison between GSH and the experiments of
\cite{Rennes,nicodemi}. (a) is the reversible branch,
final density $\rho_f$ versus relative acceleration
$\Gamma$; (b) is the irreversible ones, displaying
$\chi\equiv(\rho_f-\rho)/ (\rho_f-\rho_{cp})$ versus
the number $n$ of tapping or liquid blows. All full
lines are {\sc gsh} results. The circles in (a) and
the dotted lines in (b) are from the tapping
experiment of~\cite{Rennes}, while the dash lines in
(b) are from liquid blowing experiment and
simulation~\cite{nicodemi}. The discrepancies between
theory, experiments and simulation are clearly of the
same order of magnitude, see text for details.}
\label{fig1}
\end{figure}
When comparing experimental data to our results, there
are some obvious difficulties: \textbullet~$T_g$ is
typically nonuniform, especially when the energy is
injected from one side. Nonuniform shear flows, such
as convection, will aggravate the problem.
\textbullet~There is boundary dominance when the
system's dimension becomes a few tens of the particle
diameter. \textbullet~Crystallization may happen for
mono-disperse particles. \textbullet~The relation
between $T_g$ and the amplitude $\Gamma$ of the
perturbation needs to be evaluated for each case
individually, and is usually unknown. Moreover,
compaction experiments are executed at constant
$\Gamma$, yet we consider compaction at constant
$T_g$. As $T_g(\Gamma)$ depends in general on $\rho,
\Delta,\cdots$, the results are not equivalent.
\textbullet~Experimentally, density change is given as
a function of the number $n$ of discontinuous
perturbations such as  tapping. The theory specifies
the progress of compaction in terms of the time $t$.
The relation $t(n)$ that will depend on $\Gamma$ needs
to be specified. Given these circumstances, we shall
use some of the experimental data to extract
$T_g(\Gamma)$ and $t(n)$, and look for agreement with
the rest. This is admittedly much less than a complete
calculation, but stays true to the spirit of
accounting for the universal aspects of compaction.

Denoting the initial density as $\rho _{0}$, and the
final, $\Gamma$-dependent one as $\rho _{f}$, the
measured data of the Rennes group~\cite{Rennes} were
given as a stretched exponential (or the KWW law):
$\rho _{f}-\rho$ $=\left( \rho _{f}-\rho _{0}\right)
\exp[-\left(n/\tau _{\text{KWW}}\right) ^{\beta }]$,
with $\beta=0.6-0.8$, $\tau _{\text{KWW}}$ $=\tau
_{0}\times$ $\exp({\Gamma _{0}/\Gamma})$, $\tau
_{0}=0.9$, and $\Gamma _{0}=8$ (relative
acceleration). The Napoli group~\cite{nicodemi}
reported $\beta=1$ for their simulation and
experiments, and considered both results in agreement.

On the theoretical side, the GSH parameters are the
same we have consistently employed: $a,{\cal B}, b_0$
are as indicated above. In addition, we have $\lambda
=\lambda _{0}\left( 1-\rho /\rho _{cp}\right)$ with
$\lambda _{0}$ constant, and ${\cal
B}/P=3\times10^{6}$ (by taking the average $P$ as the
hydrostatic pressure of 0.1 m and $\alpha =0$).
Assuming an Arrhanius type law for both $T_g(\Gamma)$
and $t(n)$, we also take $T_{g}(\Gamma)=e^{-\Gamma
_{1}/\Gamma}\times \sqrt{\Lambda P/(2ab_{0}\rho
_{cp})}$ (with $\Lambda =0.28$ and $\Gamma
_{1}=0.56$), and $t/n=\tau_{n} e^{-\Gamma_{n}/\Gamma}$
(with $\Gamma _{n}=7.8$ and $\tau _{n}=2.2\times
10^{6} \sqrt{ b_{0}\rho _{cp}/{\cal B}}\,\,\lambda
_{0}^{-1}$). Employing $T_{g}(\Gamma)$, Eq~(\ref{gc6})
becomes $2\rho _{cp}/\rho _{f}=$ $1+\sqrt{1+\Lambda
e^{-2\Gamma _{1}/\Gamma}}$, which fits the
experimental data, see (a) of Fig~\ref{fig1}. Next, we
rewrite $\tau _{\rho}(\rho,T_g)$ as $\tau _{\rho}(
\rho ,\Gamma)$ using $T_{g}(\Gamma)$ and $\rho _{f}(
\Gamma)$, and numerically solve the equation $\partial
_{t}\rho =\rho /\tau _{\rho }$. The resultant
$\rho(t)\to\rho(n)$ for various $\Gamma$ is shown in
(b) of Fig~\ref{fig1}. (The assumed Arrhanius type
laws correspond to the information contained in two of
the five curves in Fig~\ref{fig1}.)

Summary: Compaction is a generic phenomenon,
indifferent to the method of perturbation. {\sc gsh}
echoes this by providing suitable mechanisms for its
two branches, yielding a transparent, de-mystified
account. This further unifies our understanding of
granular behavior as encapsulated by {\sc gsh}, which
has now been shown capable of accounting, in addition,
for static stress distribution, elastic waves,
elasto-plastic motion, the critical state, dense flow,
fluidization, and jamming.


\end{document}